%Paper: hep-th/9511204
%From: chi@hepth.cornell.edu
%Date: Tue, 28 Nov 1995 23:10:11 -0500

\input vanilla.sty
\input harvmac

\newif\iffigs\figstrue
% Uncomment next line for no figures
%\figsfalse

  \iffigs
   \input epsf
\else
   \message{No figures will be included. See TeX file for more
information}
\fi

\def\inbar{\,\vrule height1.5ex width.4pt depth0pt}
\font\cmss=cmss10 \font\cmsss=cmss8 at 8pt
\def\BZ{\relax\ifmmode\mathchoice
{\hbox{\cmss Z\kern-.4em Z}}{\hbox{\cmss Z\kern-.4em Z}}
{\lower.9pt\hbox{\cmsss Z\kern-.36em Z}}
{\lower1.2pt\hbox{\cmsss Z\kern-.36em Z}}\else{\cmss Z\kern-.4em Z}\fi}
\def\IC{\relax\hbox{$\inbar\kern-.3em\hbox{\rm C}$}}
\def\IP{\relax\hbox{\rm I\kern-.18em P}}
\def\IQ{\relax\hbox{$\inbar\kern-.3em\hbox{\rm Q}$}}
\def\IR{\relax\hbox{\rm I\kern-.18em R}}
\let\IZ\BZ
\let\ICP\IP

\def\wgt4 #1 #2 #3 #4 #5{\IP^4_{#1,#2,#3,#4,#5}}

\def\msp{\;\;\,}
\def\br{\hfill\break}

%\headline={\ifnum\pageno=1\firstheadline\else
%\ifodd\pageno\rightheadline \else\leftheadline\fi\fi}

%        \footline={\ifnum\pageno=1\firstfootline\else\otherfootline\fi}
%\def\firstfootline{\rm\hss\folio\hss}
%\def\otherfootline{\hfil}
 1
 1
 1

\font\tenbf=cmbx10
\font\tenrm=cmr10
\font\tenit=cmti10

\font\eightrm=cmr8
\font\eightit=cmti8

\parindent=1.2pc
\magnification=\magstep1
\hsize=6.0truein
\vsize=8.6truein
%\nopagenumbers

\rightline{hep-th/9511204, CLNS-95/1376}
\vglue0.3cm

\centerline{\tenbf Black Hole Condensation and the Web of Calabi-Yau Manifolds
\raise 1ex \hbox{\dag}}
\baselineskip=18pt
\centerline{\eightrm TI-MING CHIANG$^{1}$, BRIAN R. GREENE$^{1}$, MARK GROSS$^{
2}$ and YAKOV KANTER$^{1}$}
\baselineskip=18pt
\vfootnote \dag{\eightrm  Based in part on talk delivered by B.R.G at the
Trieste Duality
conference, (June 1995).}
\baselineskip=12pt
\centerline{\eightit $^{1}$ F.R. Newman Lab. of Nuclear Studies, Cornell
University}
\baselineskip=10pt
\centerline{\eightit Ithaca, NY 14853, USA}
\baselineskip=16pt
\centerline{\eightit $^{2}$ Department of Mathematics, Cornell University}
\baselineskip=10pt
\centerline{\eightit Ithaca, NY 14853, USA}

\vglue0.6cm
\centerline{\bf Abstract}
\vglue0.2cm
{\rightskip=3pc
\leftskip=3pc
\eightrm\baselineskip=10pt\noindent
We review recent work concerning topology changing phase
transitions through black hole condensation in Type II string theory.
We then also briefly describe a present study aimed at extending the
known web of interconnections between Calabi-Yau manifolds. We show,
for instance, that all 7555 Calabi-Yau hypersurfaces in weighted projective
four space are mathematically connected by extremal transitions.
\vglue0.6cm}

\tenrm\baselineskip=13pt
\leftline{\tenbf I. Introduction}
\vglue0.4cm

The distinction between classical and quantum geometry becomes most apparent
when some subspace of a classical background `shrinks down to small size'.
The precise meaning of the latter depends on details of the situation being
studied,
but for example in Calabi-Yau string compactifications the work of
\ref\rWittenphases{E. Witten, {\it Phases of N=2 Theories in Two
Dimensions}, Nucl. Phys. {\bf B403} (1993) 159.}\
\ref\rAGM{P. Aspinwall, B. Greene, and D. Morrison, {\it Calabi-Yau Moduli
Space, Mirror Manifolds, and Spacetime Topology Change in String
Theory}, Nucl. Phys. {\bf B416} (1994) 414.}\ showed markedly new quantum
geometric behaviour when certain
rational curves (two-spheres) on a Calabi-Yau shrink to zero
area. Whereas classical geometry changes discontinuously through such flop
transitions, conformal field theory changes in a perfectly smooth manner.
In this way, the above mentioned papers established that certain topology
changing transitions are physically sensible in string theory. We recall
that in these transitions, the Hodge numbers of the Calabi-Yau space do not
change; rather, more subtle topological invariants such as the intersection
form
change.

One way of summarizing the above work is to say that some points in the moduli
space of a Calabi-Yau manifold which correspond to  singular geometric
configurations actually correspond to non-singular conformal field theories.
Now, these points --- corresponding to Calabi-Yau manifolds in which
some rational curves  have degenerated --- are not the only singular
geometric configurations into which
a Calabi-Yau manifold can degenerate. A natural question, then, is whether
some
(or all) of the other kinds of degenerations are, in fact, physically sensible.
We certainly hope that all such degenerations which are at finite distance in
the moduli are physically sensible as there is no apparent mechanism
for avoiding
these theories in, for example, a cosmological setting.

For a degeneration of a Calabi-Yau manifold to a so-called `conifold' ---
something we describe in greater detail below but for now can simply be thought
of as a degeneration in which some $S^3$'s shrink to zero size --- it
was shown in \ref\rStrom{A. Strominger, {
\it Massless Black Holes and Conifolds in String
Theory}, Nucl. Phys. {\bf B451} (1995) 96}
\ that type II string theory is physically sensible
and it was shown in \ref\rGMS{B. R. Greene, D. R. Morrison, and A.
Strominger, {\it Black Hole Condenstaion and the Unification of Sting
Vacua}, Nucl. Phys. {\bf B451} (1995) 109} that certain special
kinds of conifold degenerations lead
to remarkable
 physical consequences. Namely, as in the case of the flop transitions,
these conifold degenerations provide the means for topology changing
transitions. However, in this case, even the Hodge numbers of the Calabi-Yau
jump --- in a physically sensible manner. A key difference between these
transitions
and the previous flop transitions is that the latter can be understood
perturbatively
in string theory while the former can not.

These conifold transitions thereby take us one step closer to ameliorating
the long standing vacuum degeneracy problem in string theory. Rather than each
Calabi-Yau giving us an isolated distinct vacuum, it appears that many and
possibly
all of them are linked together in a single `universal' moduli space.

In section II we will briefly recall the structure of Calabi-Yau moduli spaces
found in \rWittenphases\ and \rAGM\ with an emphasis on the geometrically
singular points. We then describe, in section III, Strominger's proposal for
making sense of the simplest kind of conifold degeneration, along the lines
of Seiberg and Witten \ref\rSW{N. Seiberg and E. Witten, {\it Electric-Magnetic
Duality, Monopole Condensation, and Confinement in $N = 2$ Supersymmetric
Yang-Mills Theory}, Nucl. Phys. {\bf B426} 19
}. In section IV we discuss the extension of
these results to the case in which the degenerating $S^3$'s are subject to
nontrivial homology relations and show that this yields the topology changing
phenomenon of {\it conifold transitions}
\rGMS. In section V we briefly discuss a study
presently underway to extend the known Calabi-Yau manifolds which are
connected to
the web through  {\it extremal} transitions.

\vglue0.6cm
\leftline{\tenbf II. N = 2 Moduli Space}
\vglue0.4cm

Here we briefly recount the phase description of N = 2 moduli space found
in \rWittenphases\ \rAGM. The reader familiar with these ideas might want
to go directly to section III.

For an N = 2 superconformal theory realized as a nonlinear sigma model
on a Calabi-Yau target space, the data necessary to specify the model is
a choice of complex structure and complexified K\"ahler class on the manifold.
The moduli space of such theories, therefore, is given in terms of the moduli
space of complex structures and complexified K\"ahler classes.
We discuss each of these in turn. Of these two, the
simplest to describe is the former, especially in the familiar context of
a Calabi-Yau manifold realized as a complete intersection in a toric variety (
such as a product of (weighted) projective spaces). For ease of discussion,
we consider a hypersurface in some weighted projective four space, although the
notions we present are well known to be far more general.
\vglue0.6cm
\leftline{\tenit a) Complex Structure Moduli Space}
\vglue0.4cm

\iffigs
\midinsert
$$\vbox{\centerline{\epsfxsize=6cm\epsfysize=4cm\epsfbox{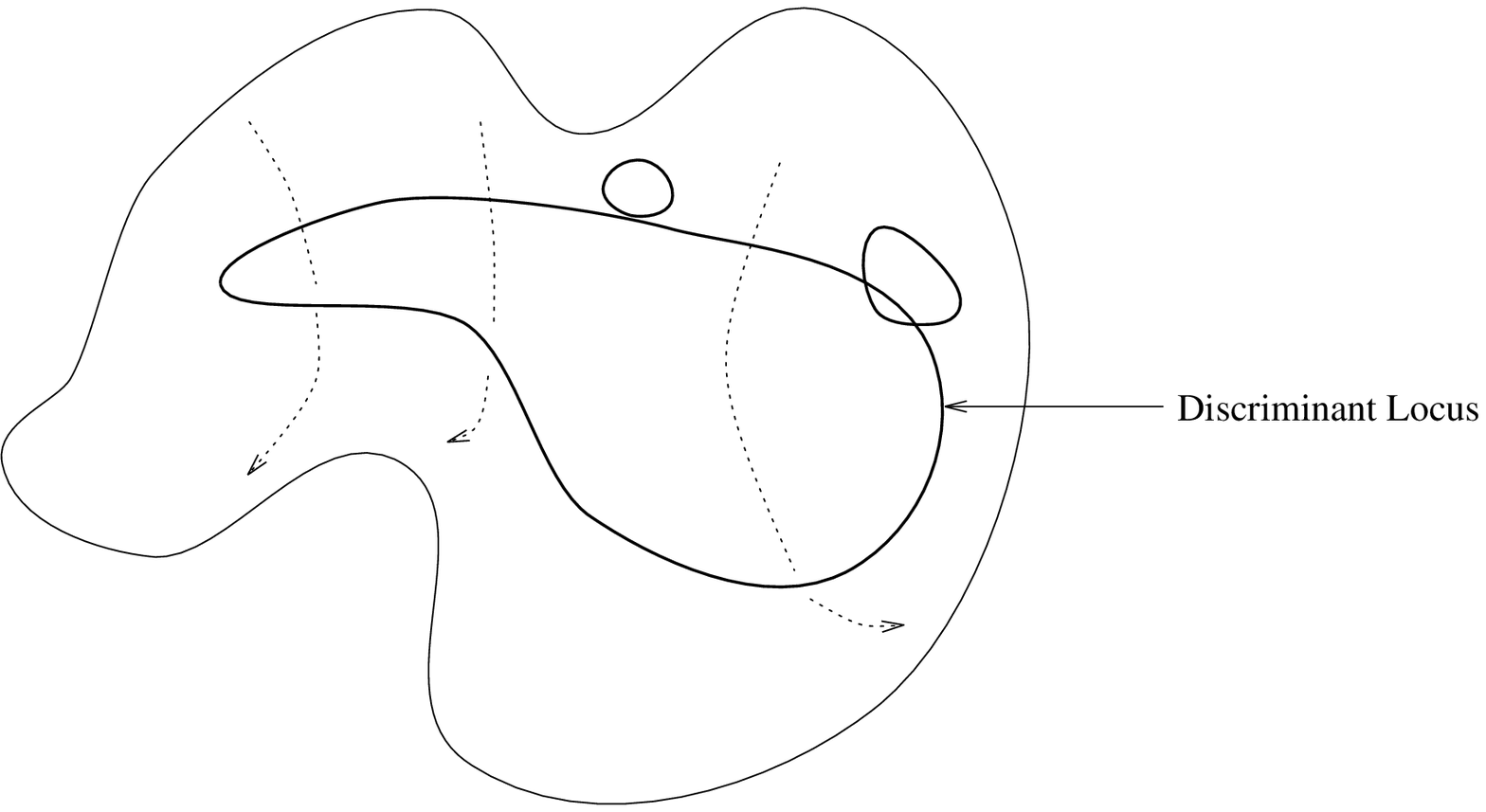}}
\centerline{Figure 1. Complex structure moduli space showing discriminant
locus.}}$$
\endinsert
\fi

The typical form of the Calabi-Yau spaces we shall consider is given by the
vanishing locus of homogeneous polynomial equations in (products of)
projective spaces.  For ease of discussion, consider the case of a single
equation $P = 0$ with
\eqn\eP{P = \sum a_{i_1...i_n} z_1^{i_1}...z_n^{i_n}.} It is known that
by varying the coefficients in such equations (modulo coordinate
redefinitions) we vary the choice of
complex structure on the the Calabi-Yau space. There is one constraint on
the choice of coefficients $a$ which we must satisfy in order
to have a classically smooth Calabi-Yau manifold: they must be chosen
so that $P$ and its partial derivatives do not have a common zero in the
defining projective space. If they did have such a common zero, the
choice of complex structure would be singular (non-transverse).
More concretely, for such  choices of complex structure
certain topological $S^3$'s on the Calabi-Yau are degenerated down to
zero size --- that is, their periods with respect to the
holomorphic three form vanish. It is
straightforward to see
that having a common zero
places one complex constraint on the
choice of coefficients and hence we can think of the space of  smooth complex
structures
 as the space of all $a$'s (modulo those which are equivalent
via coordinate transformations on the $z$'s) less this one complex
constraint. Schematically, this space can be illustrated as in figure 1
where the ``bad" choice of $a$'s is correctly denoted as the {\it
discriminant locus}. Although geometrically singular, a natural question
to ask is whether Calabi-Yau's corresponding to points on the discriminant
locus
are physically singular. We will come back to this question shortly.

\vglue0.6cm
\leftline{\tenit b) The Classical K\"ahler Moduli Space}
\vglue0.4cm

Classically, the K\"ahler form on a Calabi-Yau space is a closed two form $J$
related to the metric $g$ via
\eqn\eKahler{J = i g_{i \overline j}dX^i \wedge dX^{\overline j} .} As
such, $J$ may be thought of as an element of the vector space of all
closed two forms (modulo exact forms) $H^2(M,\IR)$. In fact, $J$ lies in
a special subspace of this vector space known as the K\"ahler cone by
virtue of its relation to the metric. In particular, since the metric
measures non-negative lengths, areas and volumes, $J$ satisfies
\eqn\eVol{\int_M J \wedge J \wedge J > 0}
\eqn\eArea{\int_S J \wedge J > 0}
\eqn\eLength{\int_C J > 0.}
where $S$ and $C$ are nontrivial 4 and 2-cycles on the
manifold respectively.

\iffigs
\midinsert
$$\vbox{\centerline{\epsfxsize=3.5cm\epsfysize=5cm\epsfbox{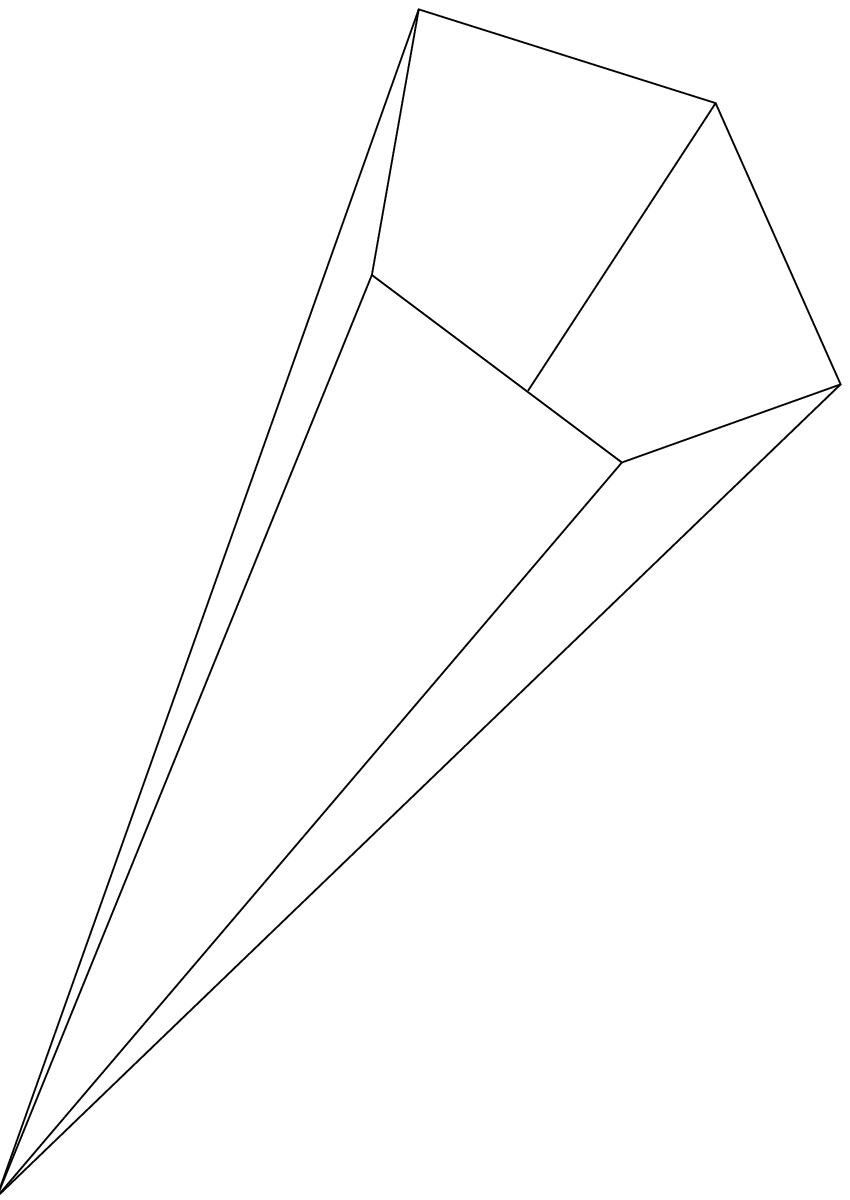}}
\centerline{Figure 2a. Schematic diagram of K\"ahler cone.}}$$
\endinsert
\fi

The space of all $J$ in $H^2(M,\IR)$ that satisfy these requirements has
a cone structure because if $J$ satisfies these conditions, so does the
positive ray generated by $J$ -- hence the name K\"ahler cone.  In figure
2a we schematically show a K\"ahler cone. A well known aspect of string
theory is that it instructs us to combine the K\"ahler form $J$ with the
antisymmetric tensor field $B$ into the complexified K\"ahler class $K =
B + iJ$.  The physical model is invariant under integral shifts of $B$
(more precisely, shifts of $B$ by elements of $H^2(M,\IZ)$) which
motivates changing variables to
\eqn\ew{ w_l = e^{2 \pi i (B_l + i J_l)} } where $(B_l, J_l)$ are
coefficients in the expansion of $B$ and $J$ with respect to an integral
basis of $H^2(M,\IZ)$. These new variables have the invariance under
integral shifts built in.

\iffigs
\midinsert
$$\vbox{\centerline{\epsfxsize=3.5cm\epsfysize=4cm\epsfbox{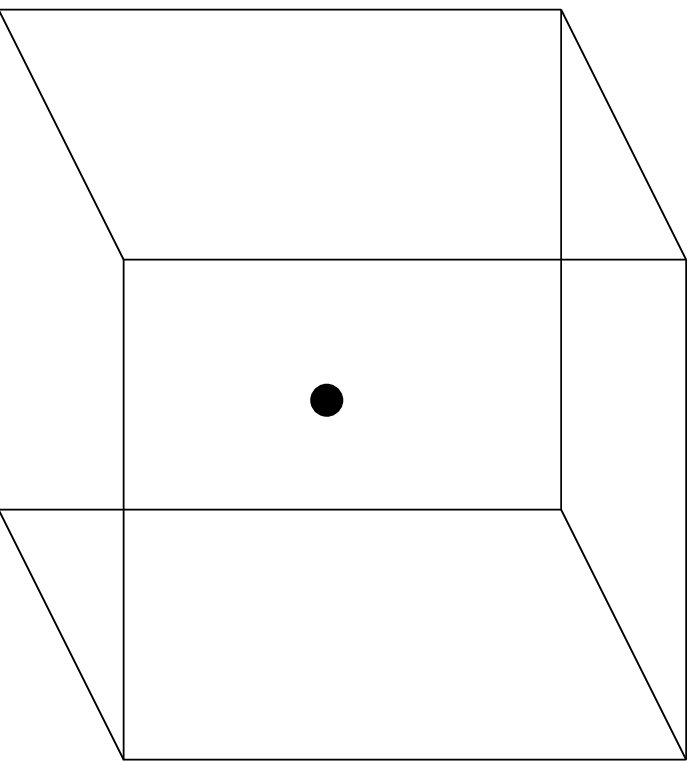}}
\centerline{Figure 2b. Complexified Kahler cone.}}$$
\endinsert
\fi

The imaginary part of $K$ satisfies the conditions on $J$ just discussed
and hence the K\"ahler cone of figure 2a becomes the bounded domain of
$H^2(M, \IC)$ in the $w$ variables as depicted in figure 2b.  We note
that the boundary of this region denotes those places in the parameter
space where the K\"ahler form $J$ degenerates in the sense that some of
the positivity requirements are violated.

\vglue0.6cm
\leftline{\tenit c) The Stringy K\"ahler Moduli Space}
\vglue0.4cm

The above description of the respective parameter spaces
led to a puzzling issue for mirror symmetry:  mirror
symmetry tells us that figure 1 and figure 2b are isomorphic if the
former is for $M$ and the latter for its mirror $W$. However, manifestly
they are not. This is not a product of our schematic drawings as there
are genuine qualitative distinctions. Most prominently, note that the
locus of geometrically singular Calabi-Yau spaces is real
codimension one in the K\"ahler parameter space, occurring on the walls
of the domain where the classical K\"ahler form degenerates. On the
contrary, the locus of geometrically singular Calabi-Yau's in the complex
structure
moduli space, as just discussed, is real codimension two (complex
codimension one). What is going on?
 The answer to this question was found
in
\rWittenphases\ and \rAGM\ and implies that:
\vskip.2in {$\bullet$ 1: Figure 2b for $W$ is only a {\it subset} of
figure 1 for $M$.  To be isomorphic to figure 1 of $M$, it must be
augmented by numerous other regions, of a similar structure, all adjoined
along common walls. This yields the {\it enlarged K\"ahler moduli space}
of $W$.}
\vskip.1in {$\bullet$ 2: Some of these additional regions are
interpretable as the complexified K\"ahler moduli space of {\it flops} of
$W$ along rational curves. In essence, to flop a rational curve (topologically
an $S^2$) we shrink it down to a point (by varying the K\"ahler structure)
and then
subsequently give it positive volume but with respect to a different topology.
This new topology is such that it agrees with the old in terms of, for
instance,
the Hodge numbers, but differs with respect to the topological intersection
form.}

\vskip.1in {$\bullet$ 3: Other regions may not have a direct sigma model
interpretation, but rather are the parameter spaces for Landau-Ginzburg
theory, Calabi-Yau orbifolds, and various relatively unfamiliar hybrid
combination conformal theories.}
\vskip.1in {$\bullet$ 4: Whereas classical reasoning suggests that
theories whose complexified K\"ahler class lies on the wall of a domain
such as that in figure 2b are ill defined, quantum conformal
field theory reasoning shows that
the generic point on such a wall corresponds to a perfectly well behaved
conformal
theory. Thus, the conformal field theory
 changes {\it smoothly} if the parameters defining it
change in a generic manner from one region to another by crossing through
such a wall. As some such regions correspond to sigma models on
topologically distinct target spaces, this last point established the
first concrete example of physically allowed spacetime topology change.}
\vskip.1in

A particularly useful way of summarizing this is as follows: classical
reasoning
suggests that our physical models will be badly behaved if the complex
structure is chosen to lie on the discrminant locus or if the K\"ahler class
is chosen to lie on a wall of the classical K\"ahler moduli space. The fully
quantum
corrected conformal field theory corresponding to such points
(yielding genus zero string theory), though, proves to be generically
{\it non-singular}
 on walls in the K\"ahler moduli space.
The pronounced distinction between the classical and stringy conclusions arises
because
such points are strongly
coupled theories (as the coupling parameter $\alpha'/R^2$ gets big as
we shrink down $R$ --- the radius of an $S^2$). Analyzing such strongly coupled
theories directly is hard; however, by mirror symmetry we know they are
equivalent
to weakly coupled field theories on the mirror Calabi-Yau space where we
can directly show them to be well behaved.

So much for the generic point on a wall in the K\"ahler parameter space:
classically they look singular but in fact they are well defined. What
about choosing the complex structure to lie on the discriminant locus (which
by mirror symmetry corresponds to a non-generic point on a wall in the
K\"ahler
parameter space of the mirror)? Might it be that these theories are well
behaved too?
At first sight the answer seems to be ``no". By taking the K\"ahler class
to be deep inside a smooth phase (i.e. a smooth large radius Calabi-Yau
background)
we trust perturbation theory and can directly compute conformal field theory
correlation functions. Some of them  diverge as we approach the
discriminant locus. This  establishes that the conformal field theory
is badly behaved. It is, however, important to distinguish between conformal
field theory and string theory. Conformal field theory is best thought of as
the effective description of string degrees of freedom which are light in the
$\lambda \rightarrow 0$ limit, with $\lambda$ being
the string coupling constant. This includes all of the familiar perturbative
string states, but effectively integrates out
  nonperturbative states whose most direct description is
in terms of solitons in the low energy effective string action
\footnote{Recently, a description of these states in terms of Dirichlet-branes
has been proposed \ref\rPolchinski{J. Polchinski, {\it Dirichlet-Branes
and Ramond-Ramond Charges}, hep-th/9510017 \br J. Polchinski, E.Witten,
{\it Evidence for Heterotic - Type I String Duality},  hep-th/9510169
\br E. Witten, {\it Bound States Of Strings And $p$-Branes}, hep-th/9510135
\br M. Bershadski, V. Sadov and C. Vafa, {\it  D-Strings on D-Manifolds}
hep-th/9510225 .}.}.

 We are thus faced with the moduli space for an
effective string description that contains points where physics appears to be
singular.
A close analog of this situation plays a central role in the celebrated work
of Seiberg and Witten \rSW\ where it is argued that the apparent singularity is
due to the appearance of new massless nonperturbative
 degrees of freedom at those singular moduli space points.
A natural guess in the present setting, then, is that the apparent singularity
encountered on the discriminant locus is due to previously
massive nonperturbative string states
becoming massless. This solution was proposed by Strominger and we review its
success in the next section.

\vglue0.6cm
\leftline{\tenbf III. Strominger's Resolution of the Conifold Singularity}
\vglue0.4cm

To quantitatively understand the proposed resolution of conifold singularities,
we must introduce coordinates on the complex structure moduli space. As is
familiar, we introduce a symplectic homology basis of $H^3(M,\BZ)$ denoted
$(A_I,B^J)$ where $I,J = 0,...,h^{2,1}(M)$ and by definition
$A_I \cap B^J = \delta_I^J$, $A_I \cap A_J = B^I \cap B^J = 0$.
We let $z^J = \int_{B^J} \Omega$ and $G_I = \int_{A^I} \Omega$, where
$\Omega$ is the holomorphically varying three-form on the family of
Calabi-Yau's
being studied. It is well known that the $z^J$ provide a good set of local
projective coordinates on the moduli space of complex structures and that
the $G_I$ can be expressed as functions of the $z^J$.

In terms of these coordinates, a conifold point in the moduli space can
roughly be
thought of as a point where some $z^J$ vanishes (we will be more precise on
this
in the next section).
The corresponding $B^J$ is called a vanishing cycle as the period of $\Omega$
over it goes to zero.
 For our purposes, there is one main implication of
the vanishing of, say, $z^J$, that we should discuss: the metric on the
moduli space
is singular at such a point. The easiest way to see this is recall that special
geometry governs these moduli spaces and therefore the K\"ahler potential
on the moduli space can be written
\eqn\eKahler{ K = -ln(i{\overline z}^IG_I - i z^I{\overline G}_I)} If we knew
the explicit form for $G_J(z)$ we would thus be able to calculate the local
form of the metric near the conifold point. Considerations of monodromy are
sufficient to do this: as we will discuss in greater generality below, if
we follow a path in the moduli space that encircles $z^J = 0$, the period $G_J$
is not single valued but rather undergoes a nontrivial monodromy transformation
\eqn\emonodromy{G_J \rightarrow G_J + z^J.}
Near $z^J = 0$ we can therefore write
\eqn\elocalform{G_J(z^J) = {{1 \over 2 \pi i}} z^J ln(z^J) + \hbox{\rm single
valued}.}
Using this form one can directly compute that the metric $g_{J \overline J}$
has a curvature singularity
at $z^J = 0$.

The reason that the singularity of the metric on the moduli space is an
important
fact is due to its appearing in the Lagrangian for the four-dimensional
effective
description of the moduli for a string model built on such a Calabi-Yau.
Namely, the nonlinear sigma model Lagrangian for the complex structure moduli
$\phi^K$ is of the form
$\int d^4x g_{I \overline J} \del \phi^I \del \phi^{\overline J} $.
Hence, when the metric on the moduli space degenerates, so apparently does our
physical description.

This circumstance --- a moduli space of theories containing points at
which physical singularities appear to develop --- is one that has been
discussed
extensively in recent work of Seiberg and Witten \rSW. The natural explanation
advanced for the physical origin of the singularities encountered is that
states which are massive at generic points in the moduli space become massless
at the singular points. As the Lagrangian description is that of an effective
field
theory in which massive degrees of freedom have been integrated out, if a
previously
massive degree of freedom becomes massless then we will be incorrectly
integrating out a massless mode and hence expect a singularity to develop.
In the case studied in  \rSW, the states that became massless were
BPS saturated magnetic
monopoles or dyons. Strominger proposed that in compactified
type IIB string theory there
are analogous electrically or magnetically charged black hole states that
become massless at conifold points. The easiest way to understand these
states is to recall that in ten-dimensional type IIB string theory there
are $3 + 1$ dimensional extremally charged
extended soliton solutions with a horizon: so called
black three-branes
\ref\rHS{G. Horowitz and A. Strominger,
 {\it Black Strings and p-Branes}, Nucl. Phys. {\bf B368} (1992) 444-462.}.
These solitons carry Ramond-Ramond charge that can be detected
by integrating the five-form field strength over a surrounding Gaussian five
cycle
$\Sigma_5$: $Q_{\Sigma_5} = \int_{\Sigma_5}F^{(5)}$. Now, our real interest is
in how this soliton appears after compactification to four dimensions via
a Calabi-Yau three fold. Upon such compactification, the three spatial
dimensions
of the black soliton can wrap around nontrivial three-cycles on the Calabi-Yau
and hence appear to a four dimensional observer as black holes states.
More precisely, they yield an $N = 2$ hypermultiplet of states. The effective
electric and magnetic charges of the black hole state are then obtained
by integrating  $F^{(5)}$ over $A_I \times S^2$ and $B^J \times S^2$.
Explicitly, making the natural assumption of charge quantization,
we can write
\eqn\echarges{
\int_{A_I \times S^2} F^{(5)} = g_5 n_I \qquad \qquad \int_{B^J \times S^2}
F^{(5)} = g_5 m_J}
where $g_5$ is five form coupling and $n_I$ and $m_J$ are integers.
 Of prime
importance is the fact that these are BPS saturated states and hence are
subject to the mass relation \ref\rCDFV{A. Ceresole, R. D' Auria, S. Ferrara
and A. Van Proeyen, {\it Duality Transformations in Supersymmetric Yang-Mills
 Theory Coupled to Supergravity}, Nucl. Phys. {\bf B444} (1995) 92.}
\eqn\eBPS{M = g_5 e^{K/2}|m^IG_I - n_Iz^I|.}
Lets consider the case in which $n_I$ = $\delta_{IJ}$ and $m^I = 0$
for all $I$ with $J$ fixed. In the conifold limit for which  $z^J$ goes to zero
we see that the mass of the corresponding electrically charged black hole
vanishes.
Hence, it is no longer consistent to exclude such states from direct
representation
in the Wilsonian effective field theory action describing the low energy
string dynamics.

The claim is that the singularity encountered above is
due precisely to such exclusion. Curing the singularity should therefore
be achieved by a simple procedure: include the black hole hypermultiplet
in the low energy effective action.
There is a simple check to test the validity
of this claim. Namely,
if we incorrectly integrate out the black hole hypermultiplet from
the Wilsonian action, we should recover the singularity discussed above.
This is not hard to do. By the structure of $N = 2$ supersymmetry, the
effective
Lagrangian is governed by a geometrical framework which is identical to that
governing Calabi-Yau moduli space. Namely, we can introduce holomorphic
projective coordinates on the moduli space of the physical model $z^J$,
and the model is determined by knowledge of holomorphic functions
$G_I(z)$. In particular, the coupling constant for the $J^{th}$ $U(1)$ is given
by $\tau_{JJ}= \partial_J G_J$. We can turn the latter statement around by
noting
that knowledge of the coupling constant effectively allows us to determine
$G_J$. We can determine the  behaviour of the coupling by a simple one-loop
Feynman diagram, which again by $N = 2$ is all we need consider. Integrating
out a black hole hypermultiplet in a neighborhood of the $z^J = 0$ conifold
point
yields the standard logarithmic contribution to the running coupling
$\tau_{JJ}$
and hence we can write
\eqn\ecoupling{ \tau_{JJ} = {{1 \over 2 \pi i}} ln z^J + \hbox{\rm single
valued.}}
{}From this we determine, by integration, that
\eqn\eperiods{ G_J = {{ 1 \over 2 \pi i}} z^J ln z^J + \hbox{\rm single
valued.}}
We note that this is precisely the same form as we found for $G_J$  earlier
via monodromy considerations. This, in fact, justifies our having referred to
them
by the same symbol. Now, by special geometry, everything about the mathematics
and physics of the system follows from knowledge of the $G_J$. For our
purposes,
therefore, the singularity encountered previously (by determining the metric
on the moduli space from the $G_J$) has been precisely reproduced by
incorrectly
integrating out the massless soliton states. This justifies the claim,
therefore,
that we have identified the physical origin of the singularity and also that
by including the black hole field in the Wilsonian action (and therefore not
making the mistake of integrating them out when they are light) we cure the
singularity.

\vglue0.6cm
\leftline{\tenbf IV. Conifold Transitions and Topology Change}
\vglue0.4cm

 In the previous section we have seen how the singularity that arises when
an $S^3$ shrinks to a point is associated with the appearance of new massless
states in the physical spectrum. By including these new massless states in
the physical model, the previous singularity is cured. In this section we
consider a simple generalization of this discussion which leads to  dramatic
new physical consequences \rGMS.  Concretely, we consider a less generic
degeneration in
which:
\vskip.2in{$\bullet$ More than one, say P, three-cycles degenerate.}
\vskip.1in{$\bullet$ These P three-cycles are not homologically independent
but rather
satisfy R homology relations.}
\vskip.1in

As we will now discuss, this generalization implies that:
\vskip.2in{$\bullet$ The bosonic potential for the scalar fields in the
hypermultiplets
that become massless at the degeneration has R flat directions.}
\vskip.1in{$\bullet$ Moving along such flat directions takes us to another
branch of
type II string moduli space corresponding to string propagation on a
topologically distinct Calabi-Yau manifold. If the original Calabi-Yau has
Hodge numbers $h^{1,1}$ and $h^{2,1}$ then the new Calabi-Yau has Hodge numbers
$h^{1,1} + R$ and $h^{2,1} - P + R$.}
\vskip.1in

In order to understand this result, there are a couple of useful pieces of
background information we should review. First, lets discuss a bit more
precisely the mathematical singularities we are considering
\ref\rLefshetzCandelas{S. Lefshetz, {\it L'Analysis Situs et la
G\'{e}om\'{e}trie Alg\'{e}braique,} Gautier-Villars, Paris, 1924;
reprinted in {\it Selected Papers,} Chelsea, New York, 1971,
p.p. 283-439, for a physicist-oriented discussion see ref \rCGH }.  As we
have discussed,
the discriminant locus denotes those points in the complex structure moduli
space of a Calabi-Yau where the space fails to be a complex manifold.
We focus on cases in which the degenerations occur at some number of isolated
points on the Calabi-Yau. In particular, we consider singularities that are
known as ``ordinary double points''. These are singular points which can
locally
be expressed in the form
\eqn\enode{\sum_{i=1}^4 w_i^2 = 0}
in $\IC^4$. This local representation is a cone with singular point at the
apex, namely the origin. To identify the base of the cone we intersect it
with a seven sphere in $\IR^8$, $\sum_{i=1}^4 |w_i|^2 = r^2$. Introducing
the complex vector
$\vec{w} = \vec{x} + i \vec{y} = (w_1,w_2,w_3,w_4)$
the equation of the intersection can be expressed as
$\vec{x} \cdot \vec{x} = r^2/2$, $\vec{y} \cdot \vec{y} = r^2/2$ and
$\vec{x} \cdot \vec{y} = 0$.
The first of these is an $S^3$, the latter two equations give an $S^2$ fibered
over the $S^3$. As there are no nontrivial such fibrations, the base of the
 cone
is $S^2 \times S^3$. Calabi-Yau's which have such isolated ordinary double
point
singularities are known as conifolds and the corresponding point in the moduli
space of the Calabi-Yau is known as a conifold point. The ordinary double point
singularity is also referred to as a node.

Having described the singularity in this way we immediately discern two
distinct
ways of resolving it: either we can replace the apex of the cone
with an $S^3$, known as a deformation of the singularity, or we can replace
the apex with an $S^2$, known as a small resolution of the singularity.
The deformation simply undoes the degeneration by re-inflating the
shrunk $S^3$ to positive size.  The small resolution, on the other hand, has
a more
pronounced effect: it repairs the singularity in a manner that changes the
topology
of the original Calabi-Yau. In essence, we shall find
the {\it physical} interpretation of these two ways of resolving conifold
singularities.

A second piece of background information is a mathematical fact concerning
monodromy. Namely, if $\gamma^a$ for $a = 1,...,k$ are $k$ vanishing
three-cycles at a conifold point in the moduli space, then another
three-cycle $\delta$ undergoes monodromy
\eqn\emonodromy{\delta \rightarrow \delta + \sum_{a=1}^k (\delta \cap
\gamma^a)
\gamma^a}
upon transport around this point in the moduli space.

With this background, we can now proceed to discuss the result quoted at the
beginning of this section. We will do so in the context of a particularly
instructive example, although it will be clear that the results are general.
We begin with the quintic hypersurface in $\ICP^4$, which is well known to
have  Hodge numbers $h^{2,1} = 101$ and $h^{1,1} = 1$. We then move to a
conifold point by deforming the complex structure to the equation
\eqn\esingular{ x_1 g(x) + x_2 h(x) = 0}
where $x$ denotes the five homogeneous $\ICP^4$ coordinates
$(x_1,...,x_5)$ and $g$ and $h$ are both generic quartics. We note that
\esingular\ and its derivative vanish at the sixteen points
\eqn\esixteen{x_1 = x_2 = g(x) = h(x) = 0.} It is straightforward to check,
by examining the second derivative matrix, that these are sixteen ordinary
double points. And, of  primary importance to our present discussion,
the sixteen singular points lie on the $\ICP^2$ contained
in  $\ICP^4$ given by $x_1 = x_2 = 0$.
This implies that
the sixteen vanishing cycles $\gamma^a, a = 1,...,16$ that degenerate to the
double points satisfy the nontrivial homology relation
\eqn\erelation{\sum \gamma^a = 0}
\ref\rCL{H. Clemens, ``Double Solids,''
Advances in Mathematics 47, 107-230, (1983).}.
We are thus in the desired situation. We proceed
with the analysis in two steps. First, we check that inclusion of the
appropriate
massless hypermultiplets cures the singularity, as it did in the simpler
case studied in \rStrom. Second, we then analyse the physical implication
of the existence of a nontrivial homology relation.
\vskip.25in
i) {\it Singularity resolution:}
\vskip.15in
We introduce a symplectic homology basis $(A_I,B^J)$ with $I,J = 1,...,204$.
By suitable change of basis we can take our sixteen vanishing cycles
$\gamma^a$ $(a=1,...,16)$ to be $B^1,...,B^{15}$ and $ -\sum_{a=1}^{15} B^a$.
As usual, we define $z^I = \int_{B^I} \Omega$ and $G_J = \int_{A_J} \Omega$.
Now, for any cycle $\delta$ we have, as discussed before, the monodromy
$\delta \rightarrow \delta + \sum_{a=1}^{16}(\delta \cap \gamma^a) \gamma^a$.
{}From this we learn that the local form of the period over $\delta$ is given
by
\eqn\eperiod{\int_{\delta} \Omega = {{1 \over 2 \pi i}} \sum_{a=1}^{16}(\delta
\cap \gamma^a)( \int_{\gamma^a} \Omega )(ln \int_{\gamma^a} \Omega) +
\hbox{\rm single valued} .}
Specializing this general expression, we therefore see
\eqn\eGperiod{ G_J = {{1 \over 2 \pi i}}(z^J ln(z^J) + (\sum_{I=1}^{15} z^I)(ln
(\sum_{I=1}^{15} z^I))) + \hbox{\rm single valued}.}
By special geometry, this latter expression determines the properties of
the singularity associated with the conifold degeneration being studied. Thus,
the question we now seek to answer is: if we incorrectly integrate out the
black
hole states which become massless at this conifold point, do we reproduce
the form \eGperiod ?

To address this issue we must identify the precise number and charges of
the states that are becoming massless at the degeneration point. As discussed
in \rStrom, the counting of black hole states is a delicate issue for which
there is as yet no rigorous algorithm. In \rStrom, one homology class in
$H^3$ degenerated at the conifold singularity and it was hypothesized that
this implies one fundamental black hole state --- the one of minimal charge ---
needs to be included in the Wilsonian action. In the present example, though,
we have sixteen three cycles in
fifteen homology classes in $H^3$ degenerating. In \rGMS\ it was argued that
this should imply sixteen fundamental black hole fields need to be included in
the Wilsonian action. Physically speaking, the black three-brane can wrap
around
any of the sixteen degenerating three-cycles, which at large overall radius
of the
Calabi-Yau would be widely separated. It thus seems sensible that even though
there
are only fifteen homology classes degenerating, we actually get sixteen
 massless
black hole states. The charges of these states are easy to derive. If we let
$H^a$ be the black hole hypermultiplet associated with the vanishing cycle
$\gamma^a$ then the charge of $H^a$ under the $I^{th}$ $U(1)$ is given by
$Q^a_I = A_I \cap \gamma^a$ where  we write $F^{(5)}$ as the self dual part
of $\sum_I \alpha^I F_I^{(2)}$ with $\alpha^I$ dual to $A_I$. We immediately
learn from this that the black holes states have charges
\eqn\echarges{Q^a_I = \delta^a_I, 1 \le a \le 15 \quad \hbox{ \rm and } \quad
Q^{16}_I = -1, 1 \le I \le 15}
with all other charges zero. This is enough data to determine the running of
the gauge couplings:
\eqn\etauij{ \tau_{IJ} = {{1 \over 2 \pi i}} \sum_{a = 1}^{16} Q^a_I Q^a_J
ln(m^a) }
where the mass $m^a$ of $H^a$ is proportional to $\sum_I Q^a_I z^I.$
Using the above charges we therefore have
\eqn\erunning{ \tau_{IJ} = {{1 \over 2 \pi i}} \delta_{IJ} ln(z^J) +
{{1 \over 2 \pi i}} ln(\sum_{k=1}^{15} z^k) + \hbox{\rm single valued.}}
Integrating we find therefore
\eqn\eGJ{ G_J = {{1 \over 2 \pi i}} z^J lnz^J +
 {{1 \over 2 \pi i}}(\sum_{k=1}^{15} z^k)(ln(\sum_{k=1}^{15} z^k))
+ \hbox{\rm single valued.}}
We note that this matches \eGperiod\ and hence we have shown that inclusion of
the sixteen black hole soliton states which become massless cures the
singularity.

Having shown that a slight variant on Strominger's original proposal is
able to cure the singularity found in this more complicated situation,
we now come to the main point of the discussion:
\vskip.25in
ii) {\it What is the physical significance of  nontrivial homology
relations between vanishing cycles?}
\vskip.15in
To address this question we consider the scalar potential governing the black
hole hypermultiplets. It can be written as
\eqn\epotential{V = \sum E^I_{\alpha \beta} E^{\alpha \beta}_I}
where
\eqn\eE{E^I_{\alpha \beta} = \sum_{a=1}^{16} Q^I_a \epsilon_{\alpha \gamma}
h^{*(a)}_{\gamma} h^{(a)}_{\beta} - (\alpha \rightarrow \beta) }
in which the indices satisfy $I = 1,...,15; \alpha, \beta, \gamma = 1,2$.
The fields $h^{(a)}_{1}$ amd $h^{(a)}_{2}$ are the two complex scalar fields
in the hypermultiplet $H^a$.

We consider the possible flat directions which this potential admits.
The most obvious flat directions are those for which
$\langle h^{(a)}_{\beta} \rangle = 0$ with nonzero values for the
scalar fields in the vector multiplets. Physically, moving along such
flat directions takes us back to the Coulomb phase in which the black hole
states are massive. Mathematically, moving along such flat directions
gives positive volume back to the degenerated $S^3$'s and hence resolves
the singularity by deformation.

The nontrivial homology relation implies that there is  another flat direction.
Since $Q^I_a = A_I \cap \gamma^a$, we see that the homology relation
$\sum_{a = 1}^{16} \gamma^a = 0 $ implies $\sum_a Q^I_a = 0$ for all $I$.
This then implies that we have another flat direction of the form
$\langle h^{(a)}_{\beta} \rangle = v^{\beta}$ for all $a$ with $v$ constant.
In fact, simply counting degrees freedom shows that this solution is unique
up to gauge equivalence. What happens if we move along this flat direction?
It is straightforward to see that this takes us to a Higgs branch in which
fifteen vectors multiplets pair up with fifteen hypermultiplets to become
massive. This leaves over one massless hypermultiplet from the original sixteen
that become massless at the conifold point. We see therefore that the spectrum
of the theory goes from $101$ vector multiplets and $1$ hypermultiplet
(ignoring the dilaton and graviphoton) to $ 101 - 15 = 86$ vector multiplets
and $1 + 1 = 2$ hypermultiplets. Now precisely these Hodge numbers arise
from performing the other means of resolving the conifold singularity (besides
the deformation) --- the small resolution described earlier! Hence, we
appear to have found the physical mechanism for affecting a small resolution
and in this manner changing the topology of the Calabi-Yau background.

Although we have focused on a specific example, it is straightforward to work
out what happens in the more general setting of $P$ isolated vanishing cycles
satisfying $R$ homology relations. Following our discussion above, we get
$P$ black hole hypermultiplets becoming massless with $R$ flat directions
in their scalar potential. Performing a generic deformation along these
flat directions causes $P - R$ vectors to pair up with the same number
of hypermultiples. Hence the Hodge numbers change according to
\eqn\eHodge{ (h^{21}, h^{11}) \rightarrow (h^{21} - (P-R), h^{11} + R).}
The Euler characteristic of the variety thus jumps by $2P$.

So, in answer to the question posed above: homology relations amongst the
vanishing cycles give rise to new flat directions in the scalar black hole
potential. Moving along such flat directions takes us smoothly to  new branches
of the type II string theory moduli space. These other branches correspond to
string propagation on
topologically distinct Calabi-Yau manifolds. We have therefore apparently
physically
realized the Calabi-Yau conifold transitions discussed some years ago ---
without
a physical mechanism --- in insightful papers of Candelas, Green and Hubsch
\ref\rCGH{P. Candelas, P.S. Green, and T. H\"ubsch, {\it Finite Distance
Between Distinct Calabi--Yau Manifolds}, Phys. Rev. Lett. {\bf 62} (1989)
1956; {\it Rolling Among Calabi--Yau Vacua}, Nucl. Phys. {\bf B330} (1990) 49
}. In the type II string moduli space with thus see that
we can smoothly go from one Calabi-Yau manifold to another by varying the
expectation values of appropriate scalar fields.

There is another aspect of these topology changing transitions which is
worthy of emphasis. In the Coulomb phase, the black hole soliton states
are massive. At the conifold point they become massless. As we move into
the Higgs phase some number of them are eaten by the Higgs mechanism with
the remainder staying massless. Now, with respect to the topology of the
new Calabi-Yau in the Higgs phase, these massless degrees of freedom are
associated
with elements of $H^{1,1}$. Such states, as is well known, are perturbative
string excitations --- commonly referred to as elementary ``particles". Thus,
a massive black hole sheds its mass, becomes massless and then re-emerges
as an elementary particle-like excitation. There is thus no invariant
 distinction
between black hole states and elementary perturbative string states: they
smoothly
transform into one another through the conifold transitions.

\vglue0.6cm
\leftline{\tenbf V. The Web of Connected Calabi-Yau Manifolds}

\vglue0.4cm

In the previous section we have seen that type II string  theory provides
us with a mechanism for physically realizing topology changing transitions
through conifold degenerations. This naturally raises two related questions:
\vskip.2in{$\bullet$ Are all Calabi-Yau manifolds interconnected through a
web of such
transitions?}
\vskip.1in{$\bullet$  Are there other kinds of singularities, besides the
ordinary double
points discussed above, which might have qualitatively different physics and
which might also have an important role in extending the Calabi-Yau web?}
\vskip.1in

        We now briefly report on work presently
being carried out which is  relevant to these two questions. A more
detailed discussion will appear elsewhere.

In an important series of papers \rCGH\ it was argued some time ago
that all Calabi-Yau manifolds realized as complete intersections in products of
(ordinary) projective space are mathematically connected through conifold
degenerations. As we mentioned above, although an intriguing prospect, it
previously seemed that string theory did not avail itself of these topology
changing transitions --- as discussed, perturbative string theory is
inconsistent at conifold points. The recent work described above shows
that inclusion of nonperturbative effects cures the physical inconsistencies,
at least in type II string theory, and hence the physical theory does allow
such
topology changing transitions to occur.

 Since the time of \rCGH, the class
of well studied Calabi-Yau manifolds has grown. Initially inspired by work of
Gepner \ref\rGepner{D. Gepner, {\it Exactly Solvable String
Compactificatios on Manifolds Of SU(N) Holonomy,} Phys. Lett. {\bf 199B},
380, (1987)  }, the class of hypersurfaces in weighted projective
four spaces has received a significant amount of attention
\ref\rmany{B. R. Greene, C. Vafa, and N. P. Warner, {\it Calabi-Yau Manifolds
 and Renormalization Group Flows} Nucl. Phys. {\bf B324} (1989) 371.\br
 E. Martinec, {\it Criticality, Catastrophes and Compactifications}
V.G. Knizhnik memorial vol. (L. Brink, et al. eds.) (1989) \br
P. Candelas, M. Lynker and R. Schimmrigk {\it Calabi-Yau Manifolds in Weighted
$\IP_4$} Nucl. Phys. {\bf B341} (1990) 383.}.
It was shown in \ref\rS{A. Klemm, R. Schimmrigk, {\it Landau-Ginsburg String
Vacua},  Nucl. Phys {\bf B411} (1994)
559,  \br  M. Kreuzer, H. Skarke,  {\it No Mirror Symmetry in Landau-Ginsburg
Spectra}, Nucl. Phys. {\bf B388} (1992)
113}\ that there are 7555 Calabi-Yau's
of this sort. Inspired by mirror symmetry, another class of Calabi-Yau's (
containing
these 7555  hypersurfaces) that have been under detailed
study are complete
intersections in toric varieties
\ref\rSomemore{L. A. Borisov, {\it Towards the mirror symmetry for Calabi-Yau
complete intersections in Gorenstein toric Fano varieties}, alg-geom/9310001.
\br
V. Batyrev and L. A. Borisov, {\it On Calabi-Yau complete intersections in
toric
varieties}, alg-geom/9412017} \rAGM\ \rWittenphases. Understanding
the structure of the moduli space of type II vacua requires that we
determine if all of these Calabi-Yau's
are interconnected through a web of  topology changing transitions.

In the following we will briefly describe a procedure for finding transitions
between Calabi-Yau manifolds realized as complete intersections in toric
varieties. The method is  elementary although at the present time
we do not have any general results on its range of applicability. Rather,
we have shown its usefulness by directly applying it to
a subclass of the Calabi-Yau's realized in this manner. We have shown, for
instance,
that all 7555 Calabi-Yau hypersurfaces in weighted projective four space
are {\it mathematically} connected to the web
\footnote{We are aware that a similar conclusion has
been reached by P. Candelas and collaborators using different
methods \ref\rPCP{P. Candelas, private communication.}.}.
We say mathematically because the
transitions our procedure yields are not all of the conifold sort. Rather,
there are Calabi-Yau's connected through more complicated singularities than
the ordinary double points used in \rGMS. For example,
some of these singularities
are such that electrically {\it and} magnetically charged black hole states
become simultaneously massless giving us an analog of the phenomenon discussed
in
\ref\rAD{P. Argyres and M. Douglas,
{\it New Phenomena in $SU(3)$ Supersymmetric Gauge Theory},
Nucl. Phys. {\bf B448} (1995) 93.}.
Arguing for physical transitions through these theories requires
more care than those involving conifold points.  Whereas the term conifold
transition refers to Calabi-Yau's linked through conifold degenerations, the
term {\it extremal} transitions
\ref\rDMLG{D. R. Morrison, ``Through the Looking Glass'', Lecture at CIRM
Conference,
Trento (June, 1994), to appear.}
 refers to analogous links through any of
the singularities encountered on the discriminant locus. At present we
only have a satisfying physical understanding of the conifold subclass of
extremal transitions.

The procedure we describe is relevant for Calabi-Yau's embedded in toric
varieties. The reader requiring background in toric geometry should consult
\rAGM\ \ref\rFulton{W. Fulton, {\it Introduction to Toric Varieties},
Annals of Math.\ Studies, vol. 131, Princeton University Press, Princeton,
1993. }.
 To keep our discussion here concise, we shall focus
on the case of hypersurfaces in weighted projective four spaces,
although we shall briefly mention some generalization
at the end of this section. As discussed in
\ref\rBatyrev{V. Batyrev,
{\it Dual Polyhedra and Mirror Symmetry for Calabi-Yau
Hypersurfaces in Toric Varieties}, J. Alg. Geom. {\bf 3} (1994) 493
},
 the data describing such Calabi-Yau manifolds is:
\vskip.2in{$\bullet$ A lattice $N \simeq \BZ^4$ and its real extension
$N_{\IR} = N \otimes_{\IZ} \IR$.}
\vskip.1in{$\bullet$ A lattice $M = Hom(N,\BZ)$ and its real extension
$M_{\IR} = M \otimes_{\IZ} \IR$.}
\vskip.1in{$\bullet$ A reflexive polyhedron $P \subset M_{\IR}$.}
\vskip.1in{$\bullet$ The dual or polar polyhedron $P^{\circ} \subset N_{\IR}$.}
\vskip.1in

Now, given the above sort of toric data for two different families of
Calabi-Yau's in two different weighted projective four spaces, how might
we perform a transition from one to the other? Well, given the polyhedra
$(P, P^{\circ})$
for one Calabi-Yau and $(Q,Q^{\circ})$ for the other, one has the  natural
manipulations of set theory  to relate them: namely, the operations
of taking intersections and unions
\footnote{The idea of trying to manipulate
the toric data of one Calabi-Yau to produce another was first
suggested by D. Morrison.
Here we present one systematic procedure, that proves
to be surprisingly robust,
for doing so.}. Consider, then, for instance, forming new toric
data by taking the intersection
$R = \hbox{\rm convex hull}((P \cap M) \cap (Q \cap M))$.
Further assume that $R$ (and its
dual $R^{\circ}$) are reflexive polyhedra
so that the singularities encountered
are at finite distance in the moduli space
\ref\rHay{Y. Hayakawa, ``Degeneration of Calabi-Yau Manifold with
Weil-Petersson
Metric'', alg-geom/9507016.}.
How are the three Calabi-Yau's
$X, Y, Z$
associated to $(P, P^{\circ})$, $(Q,Q^{\circ})$ and $(R,R^{\circ})$,
respectively, related?
The toric data contained in the polyhedron in
$M_{\IR}$ is well known to describe the complex structure deformations of
the associated Calabi-Yau realized via monomial deformations of its
defining equation
\footnote{By mirror symmetry, of course, it can also
be used to describe the K\"ahler structure on the mirror
Calabi-Yau.}.  Concretely, the lattice points in $P \cap M$
are in one-to-one correspondence with monomials in the defining equation of
$X$, and similarly for $Y$ and $Z$\ \footnote{
More precisely, some subset of these points correspond
to the toric complex structure deformations. For details
see \rAGM\ \ref\rAGMII{P. S. Aspinwall, B. R. Greene, and D. R. Morrison,
{\it The Monomial-Divisor Mirror Map},
Internat. Math. Res. Notices (1993) 319.}.}. Thus, in going from $X$ to $Z$ we
have specialized the complex structure by restricting ourselves to a subset
of the monomial deformations. This is reminiscent of the example studied  in
\rGMS, and described earlier,
in which we specialized the complex structure of the quintic from its original
$101$ dimensional moduli space to an $86$ dimensional subspace. This is not
the end of the story. Clearly the dual $R^{\circ}$ contains $P^{\circ}$.
It is also
well known that the toric data contained in the polar polyhedron describes
the K\"ahler structure deformations of the associated Calabi-Yau. Concretely,
lattice points in  $P^{\circ} \cap N$ correspond to toric divisors which are
dual to elements in $H^2(X, \BZ)$ \footnote{More precisely some subset of
the lattice points correspond to nontrivial elements in $H^2(X, \BZ)$. For
details see \rAGM\ \rAGMII.}.
Thus, in passing from $X$ to $Z$ we have
also added toric divisors, i.e. we have performed a blow-up. This again is
reminiscent of the example studied earlier: after specializing the complex
structure we performed a small resolution. All of the discussion we have just
had relating $X$ to $Z$ can be similarly applied to relate $Y$ to $Z$.
Hence, by using the toric data associated to $X$ and to $Y$ to construct
the toric data of $Z$, we have found that $Z$ provides a new Calabi-Yau that
both $X$ and $Y$ are linked to in the web.

Of course, the key assumption in the above discussion is that $(R,R^{\circ})$
provides us with toric data for a Calabi-Yau, i.e. they are reflexive
polyhedra. At present, we have not developed a general method for
picking $(P, P^{\circ})$ and  $(Q,Q^{\circ})$ such that this is necessarily
the case.
In fact, the toric data for a given Calabi-Yau is not unique but,
for instance,  depends
on certain coordinate choices. Thus the reflexivity of $(R,R^{\circ})$ or
lack thereof
depends sensitively on the coordinate choices used in representing
$(P, P^{\circ})$ and  $(Q,Q^{\circ})$. Hence, a more appropriate question is
whether there exists suitable representations of $(P, P^{\circ})$ and
$(Q,Q^{\circ})$
such that $(R,R^{\circ})$ is reflexive. In our work we have
arbitrarily chosen $(P, P^{\circ})$ and $(Q,Q^{\circ})$,
from the set of 7555 hypersurfaces, considered
a variety of coordinate representations  for each (related
by $SL(5,\BZ)$ transformations and coordinate permutations) and directly
checked to see if $(R,R^{\circ})$ is reflexive.
 If it is, then
$X$ and $Y$  are (mathematically) connected
through the Calabi-Yau $Z$. We note that, in general, $Z$ is not associated to
a Calabi-Yau hypersurface in a weighted projective space --- but rather a
Calabi-Yau
embedded in a more general toric variety.

In this manner, by
direct computer search, we have checked that all 7555 hypersurfaces in
weighted projective four space are linked (and through the process described
we have actually linked them up to numerous other Calabi-Yau's  --- the
 $Z$-type Calabi-Yau's above).
The main physical question, then, is what is the nature of the singularities
encountered when we specialize the complex structure in the manner dictated
by the intersection of $P$ and $Q$.
Analysis of the simplest examples shows that one often encounters
singularities which are qualitatively different from the
well-understood case of several ordinary double points studied in \rGMS.

To illustrate this point, and the discussion of this section more generally,
lets
consider two explicit examples.
\vskip.2in
{\it Example 1:}
\vskip.1in
Let's take $X$ to be  the family of quintic Calabi-Yau hypersurfaces
in $\ICP^4$ and $Y$ to be the family of Calabi-Yau hypersurfaces of degree 6
in
 $\wgt4 1 1 1 1 2$.
The Hodge numbers of $X$ are
$(h^{2,1}_X,h^{1,1}_X) = (101,1)$ and those of $Y$ are
$(h^{2,1}_Y,h^{1,1}_Y) = (103,1)$.
  Following the procedure described above, and recalling that
$P^{\circ} \cap N$ is given by
\eqn\etoricdatI{\matrix (1,&0, &0,& 0) \qquad & (\msp 0,&\msp 1, &\msp 0,
&\msp 0) \qquad& (\msp 0,&\msp 0,&\msp 1,&\msp 0) \cr
	(0,& 0,& 0,& 1) \qquad & (-1,&-1,&-1,&-1) \qquad& &&&& \cr \endmatrix}
and $Q^{\circ} \cap N$ by
\eqn\etoricdatII{\matrix (1,&0,&0,&0) \qquad & (\msp 0,&\msp 1,&\msp 0,&\msp 0)
 \qquad& (\msp 0,&\msp 0,&\msp 1,&\msp 0) \cr
	(0,&0,&0,&1) \qquad& (-1,&-1,&-1,&-2) \qquad& &&&&\cr\endmatrix}
we find that the toric  data for family $Z$, $R^{\circ}$, is the convex hull of
\eqn\etoricdatIII{\matrix (1,&0,&0,&0) \qquad& (\msp 0,&\msp 1,&\msp 0,&\msp 0)
 \qquad& (\msp 0,&\msp 0,&\msp 1,&\msp 0) \cr
	(0,&0,&0,&1) \qquad& (-1,&-1,&-1,&-1) \qquad& (-1,&-1,&-1,&-2) \cr
\endmatrix}
Note that for ease of presentation we are taking unions of data in $N$ space
which is dual to taking intersections in $M$ space \footnote{The duality is
only generally
valid when considering intersections and unions in $\IR^4$ instead of
$\IZ^4$.}
, discussed above.
Consider first the transition from $Y$ to $Z$. One can show that the singular
subfamily
obtained by specializing the complex structure of $Y$, in the manner discussed
above, consists of Calabi-Yau's which generically have 20 ordinary double
points all
lying on a single $\ICP^2$ and hence obeying one nontrivial homology relation.
This, therefore, is another example of the conifold transitions described in
\rGMS, reviewed in the previous sections. Thus, we can pass from $Y$ to $Z$ in
the manner discussed and the Hodge numbers change to
$(h^{2,1}_Z,h^{1,1}_Z) = (103 - 20 + 1, 1 + 1) = (84,2)$. The relation between
$X$ and $Z$, though, is more subtle. In specializing the complex structure of
$X$ dictated by the toric manipulation, we find a singular family of
Calabi-Yau's,
each generically having one singular point. The local description of this
singularity,
however, is {\it not} an ordinary double point, but rather takes the form
\eqn\eVBDP{x^2 + y^4 + z^4 + w^4 = 0.}
This singularity is characterized by Milnor number $27$ which corresponds to
$27$ homologically independent $S^3$'s simultaneously vanishing at the singular
point. Using standard methods of
singularity theory \ref\rGab{A. M. Gabrielov,  {\it Intersection
Matrixes for Certain Singularities,} Functional Analysis and its
Applications, vol. 7 No. 3 pp. 18-32} one can show that the intersection
matix of these $S^3$'s is {\it non-trivial} and has rank 20.
Mathematically, it is straightforward to show that the transition from
$X$ to $Z$ through such a degeneration causes the Hodge numbers to make the
appropriate change.

Physically, in contrast to the previous cases, not only are $A$-type cycles
shrinking down, but some dual $B$-type cycles are shrinking down as well.
{}From this we see a phenomenon akin to that studied in \rAD: we appear to have
 electrically and magnetically charged states simultaneously becoming massless
\footnote{In \ref\rWittenIII{E. Witten, {\it Some Comments on String Dynamics},
hep-th/9507121.} it was independently noted that the phenomenon of \rAD\
could be embedded in
string theory in such a manner.}.
It is such degenerations that require more care in establishing
the existence of physical transitions. This also raises the interesting
question
of whether the web of Calabi-Yau's requires such transitions for its
connectivity,
or if by following suitable paths conifold transitions would suffice.

\vskip.2in
{\it Example 2:}
\vskip.15in
We take $X$ to be the family of
quintic Calabi-Yau hypersurfaces
in $\ICP^4$ and we take $Y$ to be the family of Calabi-Yau hypersurfaces of
degree 8 in
 $\wgt4 1 1 1 1 4$. As in the previous example, the transition from $Y$ to $Z$
just involves ordinary double points, so the discussion of \rGMS\ suffices.
However, in passing from $X$ to $Z$ we encounter another type of singularity,
known as a triple point. Namely, the generic Calabi-Yau in the subfamily of $X$
obtained by specialization of the complex structure contains a single
singular point
whose local description is
\eqn\eTriple{x^3 + y^3 + z^3 + w^3 = 0.}
The Milnor number for this singularity is equal to 16, and thus in
this example we have 16 vanishing 3-cycles (homological to $S^3$'s)
simultaneously shrinking to one point. The intersection  matrix in this case
has rank 10 and we thus again are dealing with a physical situation with
massless
electrically and magnetically charged particles.
\vskip.3in

For ease, in our discussion above, we have focused on hypersurfaces in
weighted projective four space (which naturally led to hypersurfaces in
more general toric varieties). We can carry out the same program on
codimension $d$ Calabi-Yau's. As we will discuss elsewhere, for these
it is best to use the full reflexive Gorenstein cone associated with
the Calabi-Yau, but basically the idea is the same. For instance, the
union of the Gorenstein toric fan (in the N lattice) for
$\IP^6_{3,3,2,2,1,1}(5,7)$ and $\IP^6_{3,3,2,2,2,1}(5,8)$ is
Gorenstein with index 2. Hence, these codimension two Calabi-Yau's are linked
through such transitions.

In this manner we have established numerous links between Calabi-Yau's of
codimension two and between Calabi-Yau's of codimension three.
Furthermore, as in each of these classes its not hard  to construct
Calabi-Yau's
with simple toric representations of various codimension
(toric representations, of course, are not unique), we can link together
the webs of different codimension as well. For instance, the quintic
hypersurface,
which is a member of the 7555 hypersurface web, is also linked to
the web of complete intersections in products of ordinary projective spaces.
Hence,
all such Calabi-Yau's are so linked.

We therefore do not know the full answer to the two questions that
motivated the discussion of this section, but we have gained some insight
into each and hope to report on further progress shortly.

\centerline{\bf Acknowledgements}

B.R.G thanks D. Morrison and A. Strominger with whom the results we have
discussed
on conifold transitions were found.  The authors thank
P. Aspinwall,  D. Morrison and R. Plesser
for helpful discussions related to the results presented in section V.
This work was supported by the Alfred P. Sloan Foundation, a National Young
Investigator Award, and by the National Science Foundation.

\vfill
\eject

\listrefs

\bye